\documentclass[aps,prb,showpacs,floatfix,superscriptaddress,twocolumn,byrevtex,groupedaddress]{revtex4-1}
\pdfoutput=1
\usepackage{amsmath}
\usepackage{amssymb}
\usepackage{amstext}
\usepackage{amsopn}
\usepackage{amsfonts}
\usepackage{amsxtra}
\usepackage{color}
\usepackage{dcolumn}
\usepackage{graphicx}
\usepackage{bm}
\usepackage{hyperref}

\begin{document}
%
%
\title{Optical properties of Bi$_2$Te$_2$Se at ambient and high pressure}
\author{Ana Akrap}
\email{ana.akrap@unige.ch}
\author{Micha\"el Tran}
\author{Alberto Ubaldini}
\author{J\'er\'emie Teyssier}
\author{Enrico Giannini}
\author{Dirk van der Marel}
\affiliation{University of Geneva, CH-1211 Geneva 4, Switzerland}
\author{Philippe Lerch}
\affiliation{Paul Scherrer Institute, 5232 Villigen, Switzerland}
\author{Christopher C. Homes}
\email{homes@bnl.gov}
\affiliation{Condensed Matter Physics and Materials Sciences Department,
  Brookhaven National Laboratory, Upton New York, 11973, USA}
\date{\today}

\begin{abstract}
The temperature dependence of the complex optical properties of the three-dimensional
topological insulator Bi$_2$Te$_2$Se is reported for light polarized in the {\em a-b}
planes at ambient pressure, as well as the effects of pressure at room temperature.
This material displays a semiconducting character with a bulk optical gap of
$E_g \simeq 300$~meV at 295~K.
In addition to the two expected infrared-active vibrations observed
in the planes, there is additional fine structure that is attributed to either the
removal of degeneracy or the activation of Raman modes due to disorder.
A strong impurity band located at $\simeq 200$~cm$^{-1}$ is also observed.
At and just above the optical gap, several interband absorptions are found
to show a strong temperature and pressure dependence.  As the temperature
is lowered these features increase in strength and harden.  The application
of pressure leads to a very abrupt closing of the gap above 8~GPa, and
strongly modifies the interband absorptions in the mid-infrared spectral range.
While {\em ab initio} calculations fail to predict the collapse of the gap,
they do successfully describe the size of the band gap at ambient pressure,
and the magnitude and shape of the optical conductivity.
%
\end{abstract}
%
%
\pacs{72.20,-i, 74.62.Fj, 78.20.-e}

\maketitle

\section{Introduction}
A topological insulator is a material in which a large spin-orbit interaction
produces a band inversion over a bulk band gap, resulting in protected metallic
surface states.\cite{hsieh08,hasan10,moore10}  In the search for three dimensional
topological insulators, the discovery of the layered bismuth chalcogenide
Bi$_2$Te$_{3-x}$Se$_x$ family was of particular importance.\cite{zhang09,wang11}
The simple structure of the surface states in these compounds, with only one Dirac
cone traversing the band gap,\cite{hasan09,valla12} has attracted a great deal of
interest. The suitably-sized bulk band gap ($150 - 300$~meV) and robust metallic
surface states persisting up to high temperatures\cite{valla12} make this series
of compounds interesting for experimental investigation, and potentially also in
applications.
However, difficulties concerning material purity and stoichiometry lead to a
deteriorated insulating character of the bulk.  Bismuth and tellurium have very
similar electronegativity on Pauling's scale,\cite{gaudin94} and this naturally
leads to a large concentration of antisite defects. Bi$_2$Te$_3$ is not a line
compound but exists in a narrow field of compositions ($\sim 1$\%);\cite{abrikosov58}
as a result, a slight deficiency or excess of Te will result in either a {\em p}-type
or {\em n}-type material, respectively.\cite{statterthwaite57}  Very often it is
the case that Bi$_2$Te$_3$ crystals are {\em p}-type and have a high hole
concentration because of a large number of negatively charged defects due to
tellurium sites occupied by bismuth atoms.\cite{kulbachinskii01}  In
Bi$_2$Se$_3$, on the other hand, because of the high fugacity of selenium,
the concentration of selenium vacancies exceeds the concentration of
antisite defects, so that the Bi$_{2+x}$Se$_3$ crystal typically exhibits
{\em n}-type conductivity.\cite{sklenar00,plechacek02}  All of these factors
cause disorder (impurities) in the structure and lead to symmetry breaking
and intrinsic doping.  Contrary to the expected insulating state, a finite
residual conductivity appears. The salient question is how to optimize these
systems and decrease the bulk conductivity.

In an ordered stoichiometric structure Se vacancies can be decreased, and the Se/Te
randomness diminished. Recently, it was found that Bi$_2$Te$_2$Se has the highest
resistivity within the Bi$_2$Te$_{3-x}$Se$_x$ family. Shubnikov--de Haas oscillations
were observed and attributed to the surface states,\cite{ren10,jia11,ren12,xiong12}
and a single Dirac cone crosses the Fermi surface.\cite{ji12} Angle-resolved
photoemission spectroscopy (ARPES) shows narrow linewidths of topological
surface states, which indicates that the disorder is suppressed.\cite{neupane12,arakane12}

To improve existing topological insulators and design new ones, it is necessary to
understand the details of their band structure. It is important to know how the
impurities influence the conductivity, and what happens if the lattice dimensions
are varied, for example, through chemical substitutions. Much of this can be
addressed through a careful study of the optical properties. 
The optical properties of Bi$_2$Te$_3$ and Bi$_2$Se$_3$ crystals have been intensely studied in the past century.\cite{kohler74,*unkelbach73} 
More recently, this family of compounds has been revisited in light of the topological properties.\cite{laforge10,dipietro12} 
In this paper we
focus on Bi$_2$Te$_2$Se and investigate the temperature dependence of the
complex optical properties in order to improve the understanding of the bulk response.
Through room-temperature reflection and transmission experiments at ambient
and high pressure, information is obtained about the nature of the band gap,
residual conductivity, vibrational modes, and interband transitions.
The application of pressure allows the band gap to be studied as the
distance between the layers is decreased.

%
%
We observe a semiconducting response in the optical conductivity of Bi$_2$Te$_2$Se
with a gap of $E_g \simeq 2400$~cm$^{-1}$ (0.3~eV); below the gap are two strong
infrared-active phonons at $\simeq 62$ and 118~cm$^{-1}$, an impurity band
centered at $\simeq 200$~cm$^{-1}$, as well as other
weaker features.  In addition, there is evidence of a low-energy impurity
band at 30 -- 40~cm$^{-1}$, in agreement with a small transport gap
extracted from the {\em dc} conductivity measurement.  The low-energy
interband absorptions display a pronounced temperature dependence.
The application of pressure with a diamond-anvil cell in Bi$_2$Te$_2$Se
reduces the inter-layer spacing, altering the band structure. The optical
gap is at first unchanged, and then strongly reduced upon pressure increase.
Reflectance studies reveal that the band structure is altered by the application of
even a small amount of pressure.  {\em Ab initio} calculations of the
electronic structure and the optical conductivity confirm the sensitivity of interband
absorptions to $c$-axis compression. However, the calculations show that
$c$-axis compression has very little effect on the band gap.

%
%
\section{Experiment and calculation}
Single crystals of Bi$_2$Te$_2$Se were grown by the floating zone method starting
from the stoichiometric ratio of metallic bismuth and chalcogenide
elements. Cleaving the crystal exposes
lustrous mirror-like surfaces, which flake off very easily.

\begin{figure}[tb]
\centering
\includegraphics[width=1.0\columnwidth]{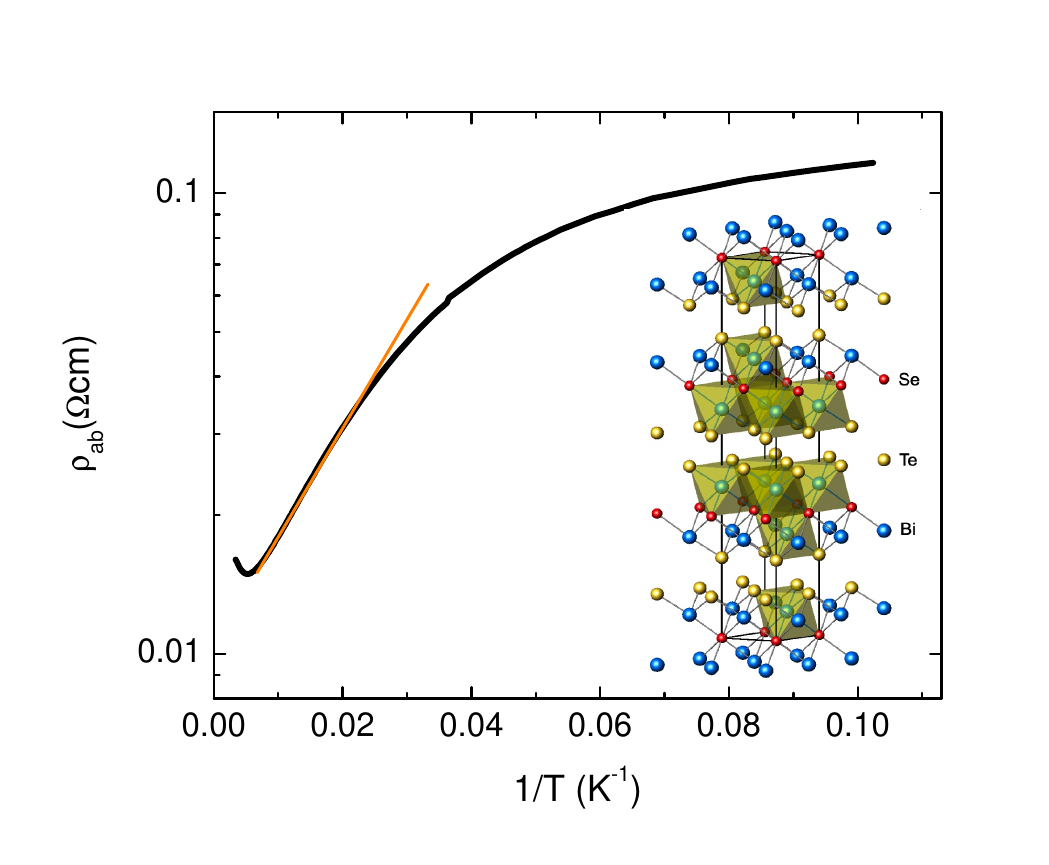}
\caption{(Color online) Arrhenius plot of the in-plane resistivity of
Bi$_2$Te$_2$Se, log $\rho$ vs $1/T$. The straight line is a fit to the
activated hopping formula, $\rho=\rho_0 \exp \left[ E_g/(k_B T)\right]$, between 30 and 170~K.
Inset: The unit cell illustrating the quintuple Te--Bi--Se--Bi--Te layers
stacked along the {\em c} axis.}
\label{fig:rho}
\end{figure}

The structure of Bi$_2$Te$_2$Se is described by the trigonal space group $R\bar{3}m$
(kawazulite) and shown in the inset of Fig.~\ref{fig:rho}.  The unit cell
consists of quintuple Te--Bi--Se--Bi--Te layers stacked along the
{\em c}-axis direction.\cite{nakajima63,bayliss91}  The quintuple layers are
bound by weak van der Waals interaction, which is responsible for the extreme
ease with which the sample may be cleaved.

The temperature-dependent reflectance at ambient pressure was measured at a
near-normal angle of incidence on a freshly-cleaved surface for light polarized
in the {\em a-b} planes, from $\simeq 12$ to over $42\,000$~cm$^{-1}$
(1.5~meV -- 5.2~eV) using an {\em in situ} evaporation technique.\cite{homes93}
The reflectance is a complex quantity, $\tilde{r}=\sqrt{R}e^{i\theta}$; however,
only the amplitude $R$ is measured in this experiment.
To obtain the phase we employ the Kramers-Kronig relation\cite{dressel-book}
using suitable extrapolations for the reflectance in the $\omega \rightarrow 0,
\infty$ limits.  At low frequencies the metallic Hagen-Rubens form for the
reflectance $R(\omega) \propto 1-\sqrt{\omega}$ is employed.  Above the
highest measured frequency point the reflectance is assumed to be constant
up to $\simeq 8 \times 10^4$~cm$^{-1}$, above which a free electron gas
asymptotic reflectance extrapolation $R(\omega) \propto 1/\omega^4$ is
assumed.\cite{wooten}

%
%
Raman spectra were collected using a home-made micro Raman spectrometer
equipped with an argon laser at a wavelength of 514.5~nm, a half-meter
monochromator and a liquid nitrogen-cooled CCD detector. The spectral
resolution was of the order of 1~cm$^{-1}$.  The sample was mounted in
a helium-flow cryostat allowing measurements down to 6~K with a long
working distance $63\times$ objective.

%
%
Optical data at high pressure were recorded using a diamond anvil cell
operated in transmission and reflection geometry.
Type IIa diamonds with a culet of 0.55~mm were
used with CuBe gaskets with a hole diameter of 250~$\mu$m to apply pressures
of up to 15~GPa inside a membrane-driven diamond anvil cell.  Samples of
Bi$_2$Te$_2$Se were cleaved to a thickness of a few microns. A fine dry KBr
powder was used as the pressure-transmitting medium, and the pressure was
monitored {\em in situ} through ruby fluorescence.\cite{datchi07}  In the
transmission geometry, light is focused on the sample in a home-made transmission
setup with a pair of reflective objectives with a magnification
coefficient of 15 and numerical aperture 0.5.  For the reflectivity measurement,
a Bruker Hyperion 3000 microscope was used with a single $15\times$ reflective objective.
Infrared radiation was provided by the Swiss synchrotron light source (SLS) and
coupled\cite{lerch12} to the Fourier transform spectrometer.   The
high-brightness provided by the synchrotron source allows a spatial
resolution of $30\times30\,\mu$m$^2$ or better while maintaining high
throughput.

%
%
The band structure was calculated for the primitive cell containing 5 atoms.
We used a full-potential linear muffin-tin-orbital (LMTO) program\cite{savrasov96}
within a local spin-density approximation (LSDA) and generalized gradient
approximation (GGA).\cite{perdew96} The effect of spin-orbit coupling was
taken into account.\cite{larson02,dai12}  The $k$-mesh was $40\times 40 \times 40$
for the self consistency calculation, and $12\times 12 \times 12$ for the calculation
of the optical conductivity. The energy cutoff was set at 730~eV, and the
convergence criterium is such that total energy accuracy is better than $10^{-6}$~eV.

%
%
%

%
%
\section{Results and discussion}
The resistivity measured in the {\em a-b} plane is shown in Fig.~\ref{fig:rho}
in an Arrhenius plot (log $\rho_{ab}$ against $1/T$). At 5~K, $\rho_{ab} \simeq
0.16\,\Omega\,{\rm cm}$.  From room temperature down to $\sim 170$~K the
behavior of the resistivity is similar to that of a bad metal. The value of
$\rho_{ab}$ at room temperature is $\simeq 16$~m$\Omega\,{\rm cm}$, and the
temperature slope is small and positive, $\simeq 1.3\times 10^{-5}$~$\Omega\,$cm/K.
Between 170 and 30~K we see activated behavior with the transport gap $E_g \simeq 5$~meV; 
a fit to the activated behavior is shown in Fig.~\ref{fig:rho}.  At the lowest temperatures the gap appears to
decrease down to $\simeq 0.43$~meV, pointing to a contribution from activation
of donor (acceptor) levels, and the metallic surface states.  In comparison with previously reported
results,\cite{ren10,jia11} our sample has a smaller activation energy;
previous reports are in the range $20 - 23$~meV.
The residual resistivity at low temperatures is $10 - 50$ times smaller than
in the previous reports.\cite{ren10,jia11}  However, the transport gaps observed
in other works in the $T\rightarrow 0$ K limit are much smaller, $\simeq 4\,\mu$eV,
hinting at a more significant contribution from activation of donor/acceptor levels 
and the surface states than in our sample. The difference in transport properties is probably
related to minute stoichiometry differences.
The measurement of thermopower on a sample from the same batch 
showed that the charge carriers in the sample are electrons.\cite{akrap12} 

%
%
\begin{figure}[tb]
\centering
\includegraphics[width=1.0\columnwidth]{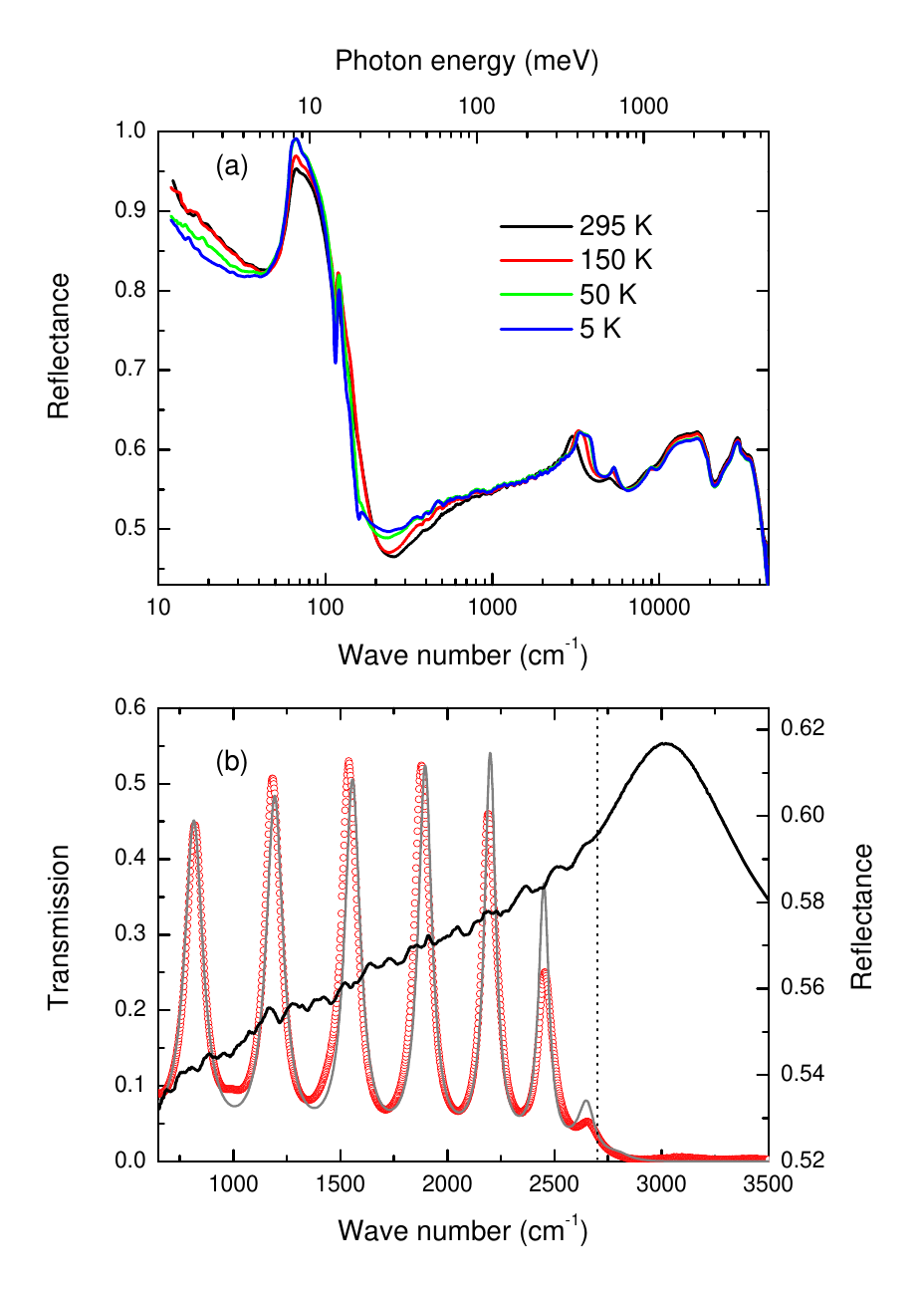}
\caption{(Color online) (a) Reflectivity of Bi$_2$Te$_2$Se over a wide
frequency range for light polarized in the {\em a-b} planes at several
temperatures.
%
(b) Transmission through a cleaved flake $\simeq 2$~$\mu$m thick
compared to the bulk reflectance at 295~K. Open red symbols are the
measured values of transmission, and the thin grey line is fit using
the Tauc-Lorentz model. The dotted line indicates the point at which
the sample becomes opaque.}
\label{fig:refl}
\end{figure}
%

%
%
\subsection{Energy gap}
The temperature dependence of the reflectance of Bi$_2$Te$_2$Se for the light
polarized in the {\em a-b} plane is shown over a wide frequency range in
Fig.~\ref{fig:refl}(a); the region close to the band edge is shown in more
detail in Fig.~\ref{fig:refl}(b) at 295~K.  The reflectance in the far infrared displays
a weakly metallic character and is dominated by a strong low-frequency phonon mode.
A weaker phonon mode can be distinguished above 100~cm$^{-1}$ as a small notch
superimposed on the stronger feature.  At higher energies structures due to the
gap edge at $E_g \simeq 300$~meV and several interband transitions are clearly
visible in the reflectance. In addition, some of these features have a pronounced
temperature dependence.
%
%

%
%
\begin{figure}[tb]
\centering
\includegraphics[width=\columnwidth]{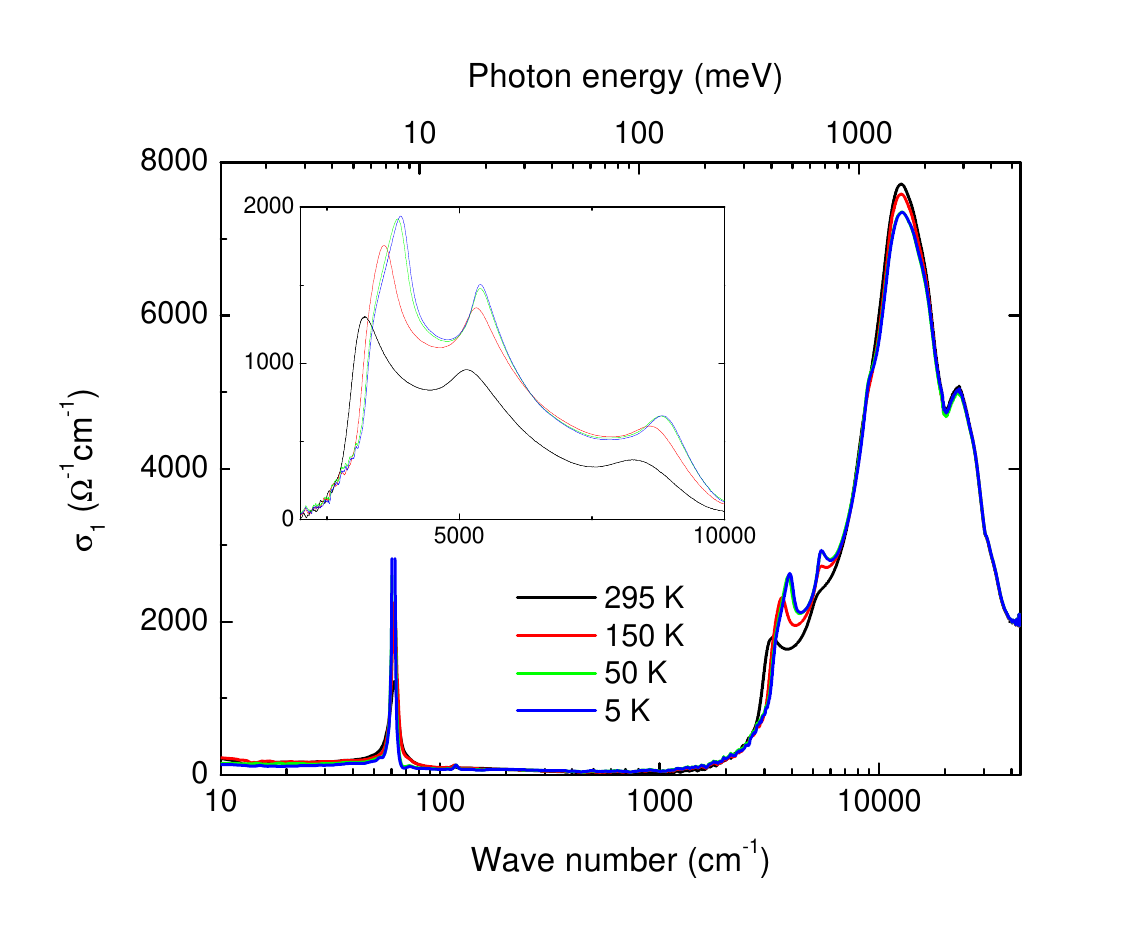}
\caption{(Color online) Optical conductivity in the planes of Bi$_2$Te$_2$Se
over a wide energy range for several different temperatures.
Inset: Optical conductivity in the mid-infrared and near-infrared regions
with the contribution from the high-frequency oscillators removed.}
\label{fig:sig1_all}
\end{figure}

The real part of the optical conductivity ($\sigma_1$) has been determined from a
Kramers-Kronig analysis of the reflectance and is shown over a wide range
in Fig.~\ref{fig:sig1_all}.   The most prominent features are the strongly
temperature-dependent interband absorptions associated with the band edge
and a very strong low-energy vibration at $\simeq 62$~cm$^{-1}$.  From
$\sigma_1$ the optical band gap can be roughly estimated to $E_g \simeq
2400$~cm$^{-1}$ (300~meV) at 295~K.  The onset of the gap in Bi$_2$Te$_2$Se
is much sharper than in Bi$_2$Se$_3$,\cite{laforge10} where the natural Se
deficiency results in intrinsic doping and a well-defined metallic (Drude-like)
contribution.
It may be noted that the transport gap extracted from the resistivity measurement
is almost two orders of magnitude smaller than the optical band gap, which suggests
that conduction occurs as a result of hopping through localized impurity states
within the gap.

%
%
An improved estimate of the band gap may be obtained from combining the
reflectance and transmission measurements. A comparison of the reflectance of
the bulk material and the transmission through a thin sample at 295~K is shown in
Fig.~\ref{fig:refl}(b).  Strong Fabry-Perot oscillations are observed in the
transmission in the mid-infrared region; however, weak fringes are also observed
in the reflectance.  The explanation for this is that the sample
cleaves extremely easily and its surface inevitably terminates in thin flakes.
A part of the incident light is reflected while another part is transmitted through
the flake and reflected from the back surface; the back reflection interferes
with the primary reflection and produces fringes. A dashed vertical line at
2700~cm$^{-1}$ (335 meV) indicates where the Fabry-Perot oscillations in both
the reflectance and transmission appear to terminate.
The transmission data has been fit using a Tauc-Lorentz model\cite{jellison96a,
*jellison96b} for the dielectric function in which a modified Lorentzian
oscillator [see Eq.~(2)] is used to describe the interband absorptions; the imaginary
part is written as
\begin{equation}
  \epsilon_{2,{\rm TL}} = \left\{
  \begin{matrix}
  \left[ \dfrac{A\, \Gamma\omega_0 (\omega-E_g/\hbar)^2}{(\omega^2-\omega_0^2)^2+\Gamma^2\omega^2} \cdot \dfrac{1}{\omega} \right], & & \hbar\omega >E_g, \\[1.0em]
  0, & & \hbar\omega \leqslant E_g. \\
  \end{matrix}
  \right.
\end{equation}
The fitted parameters $A$, $\Gamma$, $\hbar\omega_0$ and $E_g$ correspond to the
intensity, broadening, transition energy, and energy gap, respectively.
The real part is determined from a Kramers-Kronig analysis of $\epsilon_{2,{\rm TL}}$,
allowing the transmission to be calculated for a thin slab.\cite{reffit}
Pieces of varying thickness ($\simeq 0.8$ -- 9~$\mu$m) have been examined,
and the fit to the transmission through a 2~$\mu$m thick flake of Bi$_2$Te$_2$Se
is shown in Fig.~\ref{fig:refl}(b).  The band gap was determined in samples
with several different thicknesses to be $E_g \approx 285$~meV
[$E_g/(2\pi\hbar c) \approx 2300$~cm$^{-1}$], which is somewhat lower
than the energy at which the fringes are observed to disappear.

%
%
The optical properties of the bulk are described using the Drude-Lorentz model
for the complex dielectric function
\begin{equation}
  \tilde\epsilon(\omega)=\epsilon_\infty-\frac{\omega^2_{pD}}{\omega^2+i\omega/\tau_D}
          +\sum\frac{\Omega^2_j}{\omega^2_j-\omega^2-i \omega\gamma_j},
\label{eq:eps}
\end{equation}
where $\epsilon_\infty$ is the real part of the dielectric function at high
frequency, $\omega_{p,D}^2 = 4\pi ne^2/m^\ast$ and $1/\tau_D$ are the square
of the plasma frequency and scattering rate for the delocalized (Drude) carriers,
respectively, and $m^\ast$ is an effective mass.  For the Lorentz
oscillators $\omega_j$, $\gamma_j$ and $\Omega_j$ represent the position,
width, and strength of the $j$th vibration or excitation.  The complex
conductivity is $\tilde\sigma=\sigma_1+i\sigma_2= -i\omega
[\tilde\epsilon(\omega)-\epsilon_\infty]/4\pi$.
%
%
\begin{figure}[tb]
\includegraphics[width=\columnwidth]{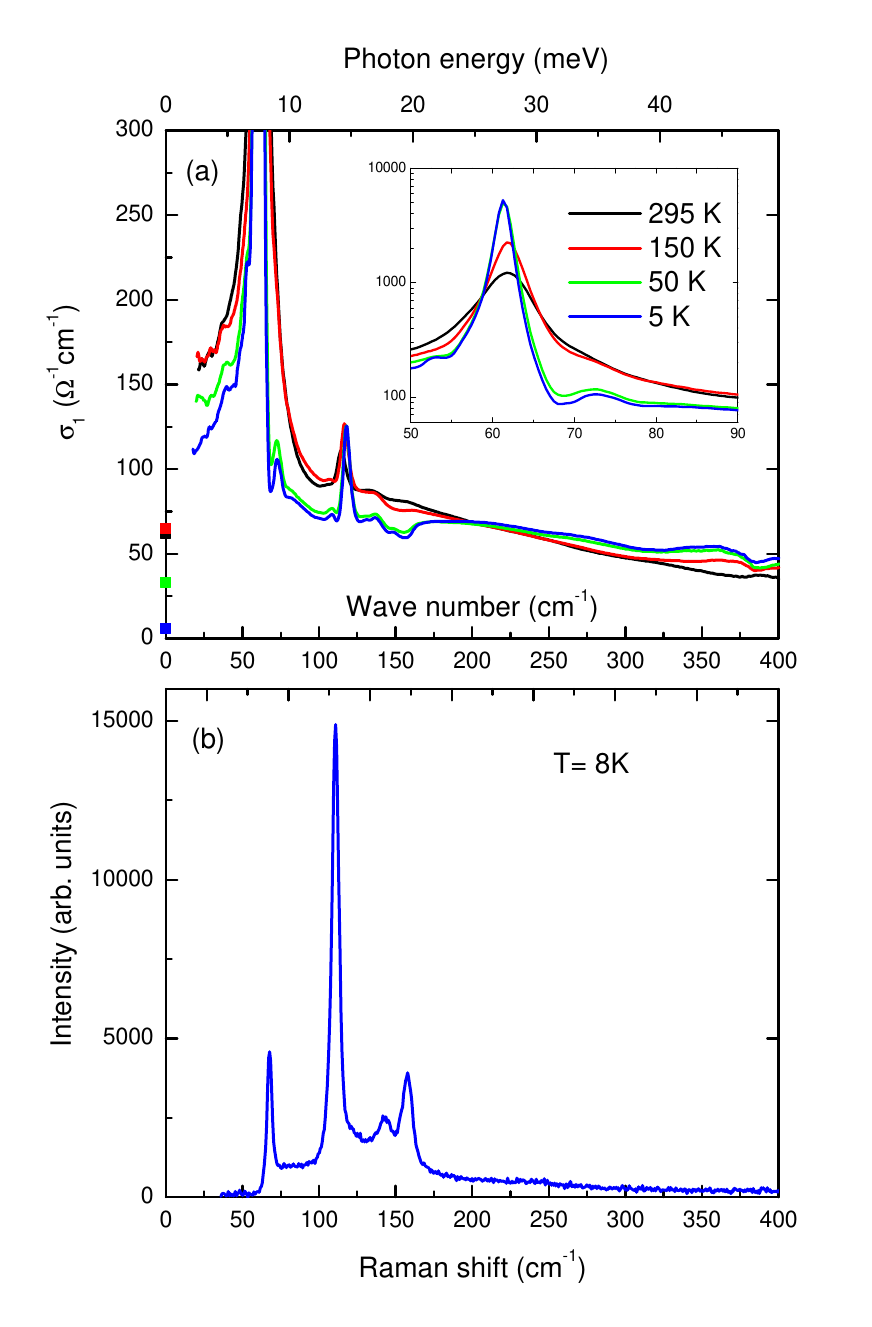}
\caption{(Color online) (a) The temperature dependence of the real part of
the optical conductivity of Bi$_2$Te$_2$Se for light polarized in the planes
at low energies. The symbols at the origin denote the transport values for
the {\em dc} conductivity. Inset: The conductivity in the
region of the strong optical phonon showing the asymmetry in the feature at
low temperature.
(b) The Raman response at 8~K.}
\label{fig:sig1_FIR}
\end{figure}
In addition to the vibrational parameters, discussed in Sec.~III~B, a fit
to the low-frequency conductivity at room temperature yields $\omega_{pD}
\simeq 1300$~cm$^{-1}$ and $1/\tau_D \simeq 200$~cm$^{-1}$; this
value for the plasma frequency corresponds to a carrier concentration of
$n/m^\ast \simeq 1.7 \times 10^{18}$~cm$^{-3}$, consistent with
previous results.\cite{statterthwaite57}

%
%
One of the more interesting aspects of this material is the unusual
temperature dependence of the interband transition associated with the
gap edge, shown in Fig.~\ref{fig:sig1_all}, in which the increase in
intensity, or spectral weight (defined as the area under the conductivity
curve over a given frequency interval) of this feature, or set of
features, in the $2500 - 7000$~cm$^{-1}$ region, is compensated for by
a loss of spectral weight at high frequency.  This behavior may be shown
in more detail by fitting the strong high-frequency features at $\simeq
13\,000$ and $24\,000$~cm$^{-1}$ and then subtracting this contribution from
the conductivity. The result of this procedure is shown in the inset of
Fig.~\ref{fig:sig1_all}.  At low temperature the peak of the lowest interband
absorption hardens by $\simeq 700$~cm$^{-1}$ and develops a weak shoulder,
while the two higher-energy features appear symmetric and display smaller
blue shifts.  The gap edge appears to harden by $\simeq 430$~cm$^{-1}$.

%
%
The thermal shifts in the gap edge and the absorption peaks are significant.
One effect of the decrease in temperature is that the lattice contracts, which
leads to an increase in the band gap.\cite{lautenschlager85} However, a more
important process is the phonon absorption and re-emission. The highest-energy
phonons in the system (Sec.~III B) set a temperature scale of approximately 170~K.
When the temperature is much lower than this, the phonon processes become inaccessible
and the gap increases.  While the strong temperature dependence of the gap
may be understood with phonons, the unusual shifts of the interband absorptions
are more likely linked to the contraction of the lattice.

%
%
At long wavelengths the optical conductivity is consistent with the
{\em dc} transport: the conductivity is very small and it further
decreases as temperature is lowered.  As shown in Fig.~\ref{fig:sig1_FIR}(a),
the values of $\sigma_1$ extrapolate roughly to the values of the {\em dc}
conductivity.  Since the {\em dc} conductivity at low temperatures is finite,
there is also a weak free-carrier (Drude-like) contribution.  However, it is
impossible to  accurately model this since it is masked by larger effects;
a strong background which peaks around the 62~cm$^{-1}$ phonon and persists
up to at least 400~cm$^{-1}$.
A part of this broad background may be identified as a far-infrared band due
to the presence of impurities. The impurity band is slightly enhanced as
the temperature is lowered.  This is in agreement to what was previously
reported in Bi$_2$Te$_2$Se and attributed to impurities.\cite{dipietro12}
A similar impurity band was also observed several years ago in a series of doped
semiconductors, Si:P.\cite{gaymann95}  The presence of impurity states in the
low-energy excitations agrees with the small transport gap obtained from the
resistivity measurement.  A wide impurity band is centered at $200 - 300$~cm$^{-1}$,
but there may be another impurity band centered at $30 - 40$~cm$^{-1}$
($\sim 4 - 5$~meV), lying in the side band of the strong optical phonon.
While the latter impurity band would agree with the small activation energy 
obtained in the resistivity measurement, it is possible that this feature is instead 
linked to the strong low-lying $E_u$ phonon, discussed in the following Section.

%
%
\subsection{Vibrational properties}
The optical conductivity of Bi$_2$Te$_2$Se allows the vibrational properties to
be analyzed.  As Fig.~\ref{fig:sig1_FIR}(a) shows, only two sharp phonon peaks are
observed at room temperature. At low temperatures these vibrations further sharpen, but
in addition several other, albeit much weaker, modes appear. The symmetry group
analysis of Bi$_2$Te$_2$Se in the $R\bar{3}m$ setting gives the irreducible
vibrational representation\cite{kohler74,richter77}
\begin{equation}
  \Gamma_{vib}=2A_{1g}+2E_g+2A_{2u}+2E_u .
\label{eq:vib}
\end{equation}
Here, the $A_{1g}$ and $E_g$ modes are Raman-active. The $A_{2u}$ and $E_u$
modes are infrared active along the {\em c} axis and the {\em a-b} planes,
respectively.  The two $E_u$ modes can be identified as the two strong vibrations
observed at 62~cm$^{-1}$ and 117~cm$^{-1}$.\cite{cheng11}
As the temperature is  decreased, the low-energy mode softens, while the
higher-energy mode hardens.  At low temperatures the 62~cm$^{-1}$ mode,
shown in the inset of Fig.~\ref{fig:sig1_FIR}(a), appears to develop a slight
asymmetry, with a tail at low frequencies. An inhomogeneous structure would be
enough to cause an asymmetry in such a strong phonon mode; however,
small level shifts in the reflectance can also artificially produce an asymmetry
in the line shape of a strong phonon in an insulator.

%
%
\begin{table}
\caption{The fitted parameters for the infrared-active modes
in the optical conductivity at 5~K and the Raman-active modes at
8~K.  The estimated errors are indicated in parenthesis.
With the exception of the Raman intensities, all units are
cm$^{-1}$.}
\begin{ruledtabular}
\begin{tabular}{c c c c}
\multicolumn{4}{c}{Infrared$^{\rm a}$} \\
  mode & $\omega_j$ & $\gamma_j$ & $\Omega_j$ \\
         &  51.6  (0.1) &  8.3 (0.7)  & 133 (11) \\
   $E_u$ &  61.7  (0.03) &  2.0 (0.1)  & 936 (25) \\
         &  72.7  (0.04) &  7.0 (0.3)  & 114 (4) \\
         & 108.0  (0.1) &  6.1 (0.7)  &  40 (4)\\
   $E_u$ & 117.8  (0.02) &  6.0 (0.1)  & 159 (2)\\
         & 137.1  (0.08) &  8.5 (0.5)  &  55 (2)\\
         & 149.0  (0.09) &  5.0 (0.5)  &  22 (2)\\
\multicolumn{4}{c}{(Antiresonance$^{\rm b}$)} \\
         & 144.1 (0.1)   &  9.0 (0.5)  &  68 (3) \\
         & 155.8 (0.1)   & 13.1 (0.8)  & 100 (4) \\

%
 & & & \\
\multicolumn{4}{c}{Raman} \\
  mode & $\omega_j$ & $\gamma_j$ & $I/I_0$ \\
   $A_{1g}/E_g$ &  67.7 (0.1)  &  3.5 (0.2)  & 0.21 \\
   $A_{1g}/E_g$ & 110.7 (0.1)  &  5.1 (0.1)  & 1.00 \\
   $A_{1g}/E_g$ & 142.2 (0.5)  & 19.6 (3.5)  & 0.35 \\
   $A_{1g}/E_g$ & 157.7 (0.2)  &  7.1 (0.4)  & 0.27 \\
\end{tabular}
\end{ruledtabular}
\footnotetext[1] {Two very weak modes at $\approx 39$ and 85~cm$^{-1}$ are
  identified but not fit.}
\footnotetext[2] {The modes at 137.1 and 149~cm$^{-1}$ are fit using antiresonances.}
 \label{tab:phonons}
\end{table}

The infrared-active modes observed at 5~K have been fit to Lorentz oscillators
with a polynomial background and the frequencies, widths and strengths are
listed in Table~\ref{tab:phonons}. Contrary to previous experiments on
Bi$_2$Se$_3$,\cite{laforge10} we can find no convincing evidence of
Fano-like behavior in the strong phonon at $\simeq 62$~cm$^{-1}$.
In addition to the two allowed in-plane infrared modes, there are at
least seven other infrared-active modes observed.  It is possible that
disorder may lead to the lifting of the degeneracy of the $E_u$ modes,
leading to weak splitting.  However, we note that the fit to the conductivity
by employing two resonances at $\simeq 137$ and 149~cm$^{-1}$ may actually be
improved by considering two Fano antiresonances\cite{homes95} at $\simeq 144.1$
and 155.8~cm$^{-1}$, respectively.
The Raman spectrum is shown at 8~K in Fig.~\ref{fig:sig1_FIR}(b).  The energies of
the Raman-active modes are very close to the energies of some of the weak phonons
that appear in the conductivity at low temperature; it is possible that symmetry
breaking may lead to the activation of Raman modes that are normally not infrared
active. In particular the positions of the two antiresonances in the conductivity
are very close to the two high-frequency Raman modes at 142 and 158~cm$^{-1}$
(Table~\ref{tab:phonons}).  In low-dimensional materials, the totally-symmetric
$A_{1g}$ Raman modes have been observed to undergo out-of-phase coupling to produce
an optically-active mode at the same frequency as the Raman mode;\cite{rice76} in
systems where these modes sit upon a strong electronic background they appear
as antiresonances rather than resonances.\cite{jacobsen83}  A similar mechanism
may be responsible for the Fano-like line shapes observed in this material.
%
%
Since we seem to observe at least nine modes in total, it is also possible that
the $A_{2u}$ out-of-plane modes may also be activated by disorder or a slight
misalignment; these vibrations typically manifest themselves in the planes at the
longitudinal-optic positions.\cite{reedyk93}  Overall, the low-energy optical and
vibrational properties of Bi$_2$Te$_2$Se point towards the importance of substitutional
disorder and intrinsic doping.

%
%
\subsection{Pressure}
The structure of Bi$_2$Te$_2$Se appears to be ``soft'' or easily compressible
along the {\em c}-axis, where the quintuple atomic layers are linked by weak
van der Waals interaction (inset of Fig.~\ref{fig:rho}).  A natural question is
how sensitive this material is to pressure. Indeed, the changes in the gap under
pressure in the related compounds Bi$_2$Te$_3$ and Bi$_2$Se$_3$ suggest that
hydrostatic pressure is very effective at tuning the band structure.
In Bi$_2$Te$_3$, Vilaplana {\em et al.}\cite{vilaplana11a,*vilaplana11b}
report reducing the gap from 170 to 120~meV by applying 6~GPa.
In Bi$_2$Se$_3$ the optical gap is increased from 170~meV at ambient pressure
to 450~meV at 8~GPa.\cite{segura12}

%
%
\begin{figure}[bt]
\centering
\includegraphics[width=\columnwidth]{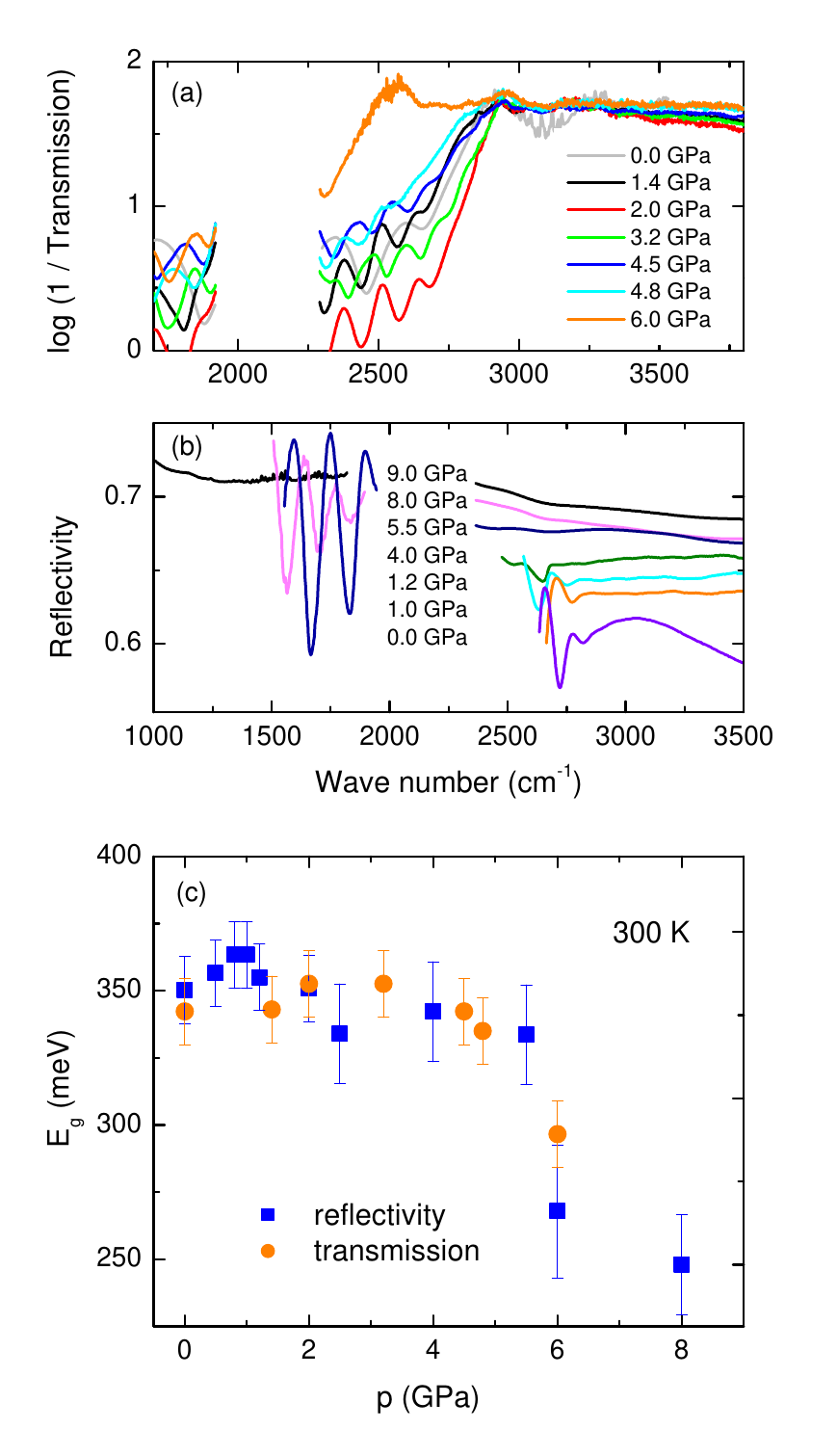}
\caption{(Color online) The frequency dependence at room temperature
of (a) the log of the inverse transmission and (b) the reflectivity,
for a thin flake of Bi$_2$Te$_2$Se at various pressures. The spectral
range between 2000 -- 2300~cm$^{-1}$ is not shown due to the strong
absorptions of the diamond in this region.
(c) The comparison of the pressure dependence of the bulk band
gap $E_g$ determined from reflectivity and transmission measurements.}
\label{fig:T_p}
\end{figure}

In Fig.~\ref{fig:T_p}(a), $\log (1/t)$  is plotted as a function of
frequency for different pressures at room temperature. Transmission
through a thin film of thickness $d$ is given by\cite{vanheumen07}
\begin{equation}
  t(\omega) \approx \frac{1}{1+4\pi d\sigma_1(\omega)/c}
\end{equation}
so that $1/t$ represents an approximation to the real part of the conductivity,
offset by a constant.  The gap may be simply estimated as the energy at half of
the maximum value of $1/t(\omega)$.
The oscillations below the gap edge are caused by the Fabry-Perot interference,
previously illustrated in Fig.~\ref{fig:refl}(b).
The pressure dependence of the gap may also be obtained from the reflectivity
measurements, shown in Fig.~\ref{fig:T_p}(b). Below the gap edge, strong Fabry-Perot
oscillations appear and the energy of their onset may be taken as the value of the gap.
Fig.~\ref{fig:T_p}(c) shows the pressure dependence of the gap determined
from the transmission and reflectivity measurements as described above;
the two estimates agree very well.

Up to 4~GPa, the effect of pressure on the gap edge in Bi$_2$Te$_2$Se is fairly
modest, especially when compared to Bi$_2$Te$_3$ or
Bi$_2$Se$_3$.\cite{vilaplana11a,segura12}  However, a steep decrease in
the gap value begins above 4~GPa, and by 8~GPa the gap has moved
from $\sim 340$ to $\sim 250$~meV.
%
%
\begin{figure}[tb]
\centering
\includegraphics[width=0.9\columnwidth]{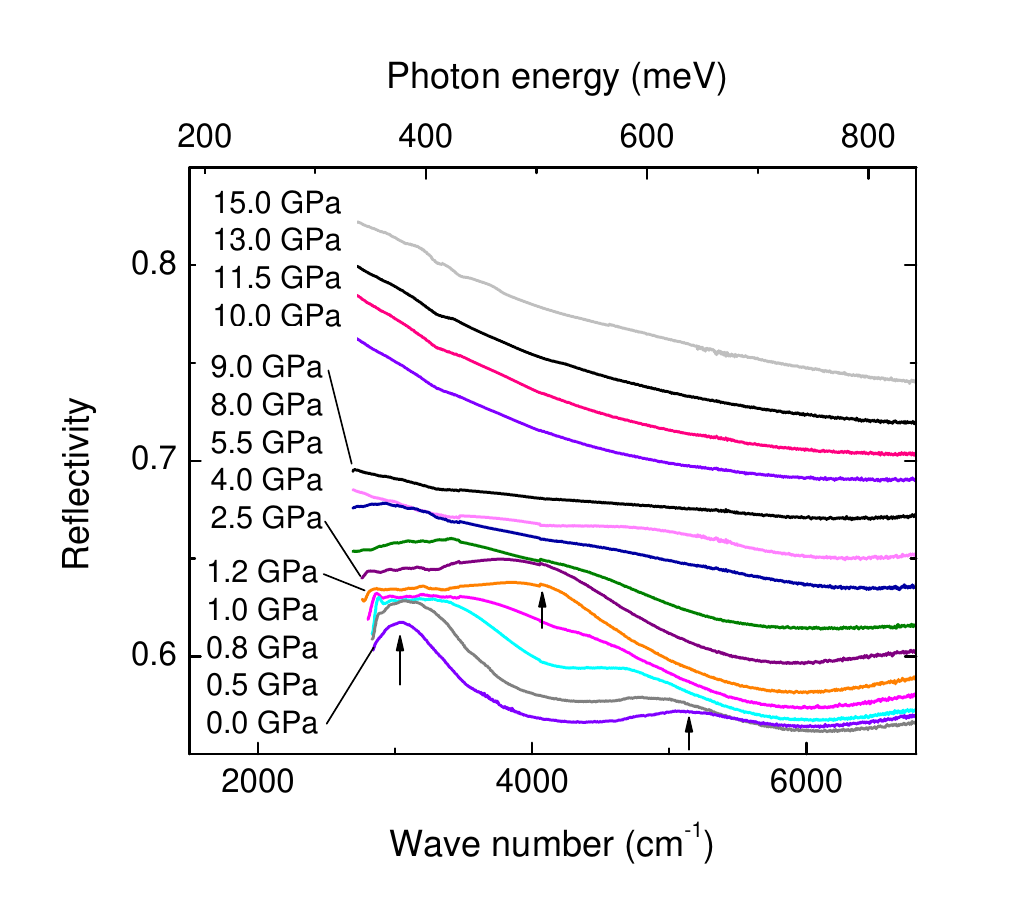}
\caption{(Color online) The detailed pressure dependence of the reflectivity
of Bi$_2$Te$_2$Se at room temperature.  The arrows indicate the location of
features associated with interband absorptions.}
\label{fig:R_p}
\end{figure}
A wider spectral range for the reflectivity measured under pressure is shown
in Fig.~\ref{fig:R_p}. For clarity, only the data above $\sim 2800$~cm$^{-1}$ is shown.
Already below 1~GPa, the reflectivity changes significantly as pressure is applied.
At the lowest pressures, up to 0.8~GPa, the two lowest interband transitions at
$\simeq 3000$ and 5000~cm$^{-1}$ are still distinguishable. As the pressure increases
to 1.0~GPa, these two peaks merge into a single peak at $\simeq 4000$~cm$^{-1}$.
When the pressure is further increased, the reflectivity level increases.
Above $p \simeq 9$ -- 10~GPa the gap can no longer be followed.
Extrapolating the data in Fig.~\ref{fig:T_p} the system is expected to become metallic
above 10~GPa.

\begin{figure}[tb]
\centering
\includegraphics[width=\columnwidth]{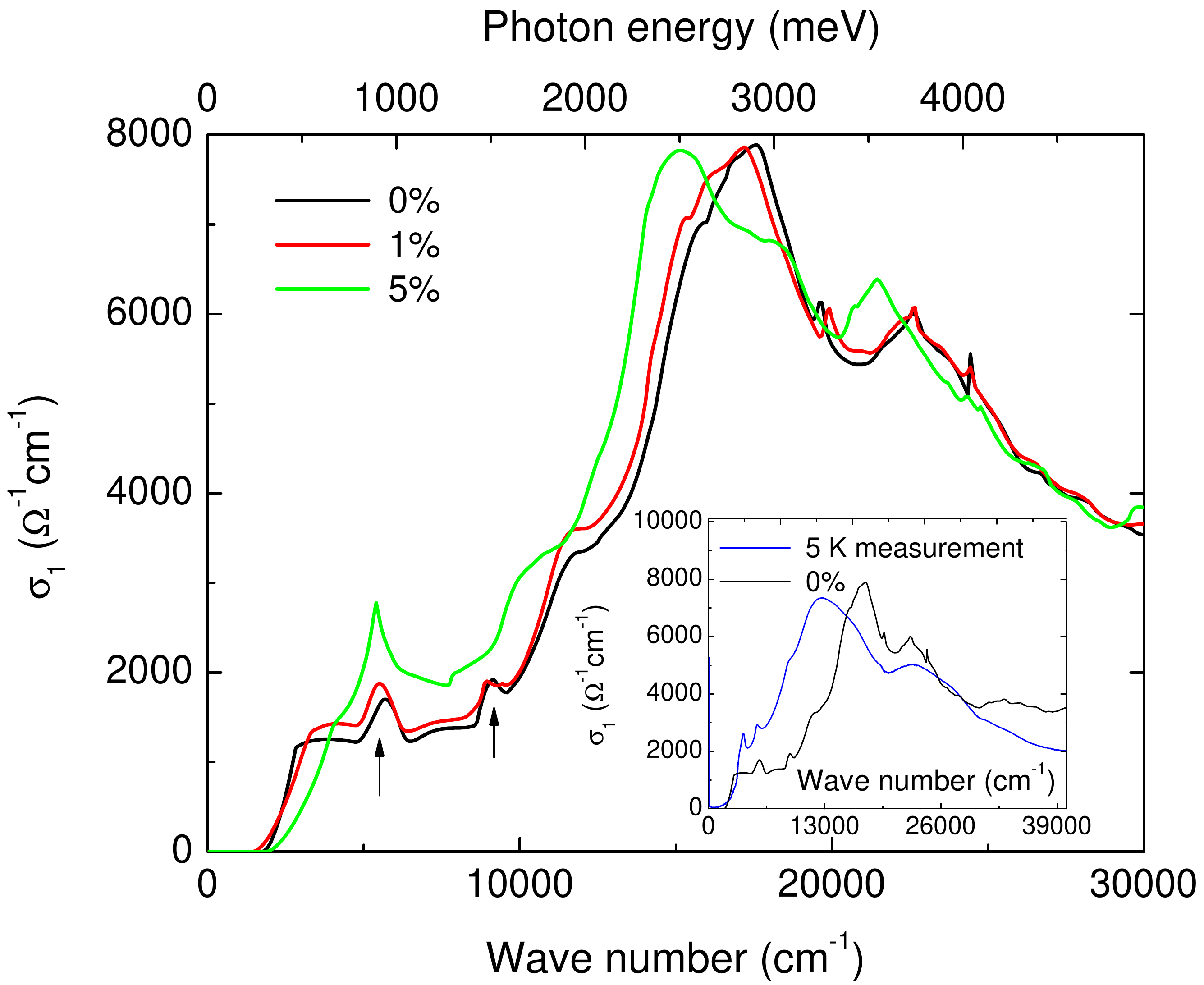}
\caption{(Color online) First principles calculations (including spin-orbit coupling)
of the real part of conductivity, $\sigma_1$, for various lattice compressions along
the {\em c} axis.
Inset: Comparison of calculated $\sigma_1$ and the measured optical conductivity at
5~K at ambient pressure.}
\label{fig:LDA}
\end{figure}

In order to compare our data with the band structure predictions, the inset of
Fig.~\ref{fig:LDA} shows a comparison between the experimentally determined
$\sigma_1$ at 5~K and the calculated $\sigma_1$ for an uncompressed lattice.
Overall, the agreement is quite good and the value of the gap is approximately
correct. The absorption bands are all reproduced, despite a minor shift in
frequency.

The main panel of Fig.~\ref{fig:LDA} shows the calculated conductivity $\sigma_1$
for different lattice compressions. The effect of pressure on this layered
structure was simulated as a reduction of inter-planar distances along the {\em c}
axis while keeping the {\em a-b} plane lattice parameters constant.
In this way we only change the distance between quintuple layers, linked by
van der Waals interactions. All the other bonds are covalent and therefore much less
compressible.
The calculated gap edge does not significantly depend on $c$-axis compression.
On the contrary, the absorptions above the gap edge, marked by arrows in
Fig.~\ref{fig:LDA}, show pressure dependence. For 5\% compression of the {\em c} axis,
only one peak remains below $10\,000$~cm$^{-1}$.  This agrees with the reflectivity
under pressure, which shows that the two structures below 6000~cm$^{-1}$ merge into
one for pressures above 1.0~GPa.
%
%

Overall, the high-pressure behavior of optical properties in Bi$_2$Te$_2$Se suggests
that its band structure is very sensitive to pressure. The collapse of the gap does
not seem to be captured by the LDA calculations in which only $c$-axis compression
is considered. This indicates that the changes of the geometry within a quintuple layer
may play an important role in the closing of the gap under pressure.

%
%
\section{Conclusions}
The optical properties of Bi$_2$Te$_2$Se have been determined over a wide temperature
and frequency range.  The far-infrared optical conductivity indicates the presence of
an impurity band at $\simeq 200$~cm$^{-1}$ (25~meV), well below the band gap. Numerous
in-plane vibrational modes can be identified at 5~K in Bi$_2$Te$_2$Se, at least seven
more than what is allowed by symmetry. Both the extra phonon modes and the impurity
band are consistent with presence (and importance) of disorder and symmetry breaking
in this material.

The band gap edge is determined to be $E_g \simeq 300$~meV at room temperature.
At low temperatures $E_g$ shifts strongly to higher energies.  Similarly, several
interband transitions or absorptions above the gap edge show strong temperature
dependence. Their behavior indicates that the band structure is significantly
influenced by thermal contraction of the lattice.

The high-pressure optical properties of Bi$_2$Te$_2$Se have been
determined at room temperature. A strong suppression of the gap edge occurs
above 4.5~GPa.  Commencing at low pressures, $p<1.0$~GPa, absorptions above the
gap edge show a strong pressure dependence. Our measurements are in reasonable
agreement with LSDA calculations.

%
\begin{acknowledgements}
We would like to thank Philip Allen and Alexey Kuzmenko for helpful discussions.
Thanks are due to Alfredo Segura for very useful technical advice.  Research was
supported by the Swiss NSF through grant No. 200020-135085 and its NCCR MaNEP.
Work at BNL was supported by the U.S. Department of Energy, Office of Basic Energy Sciences,
Division of Materials Sciences and Engineering under Contract No. DE-AC02-98CH10886.
A.A. acknowledges funding from ``Boursi\`eres d'Excellence''
of the University of Geneva.
\end{acknowledgements}
\vfil
\bibliography{bi2te2se}
\end{document}